# Pairing with non-opposite momentum in superconductors

Tian De Cao

*Department of physics, Nanjing University of Information Science & Technology, Nanjing 210044, China*

The basic feature of the BCS theory is that pairing occurs between electrons in states with opposite momentum and opposite spins, e.g., between states (k,↑ ) and (-k,↓), in which the symbols of vectors are neglected, and later theories followed the BCS theory. However, since the center of mass for any pair is not always static, we must consider the pairing between states (k±q,↑ ) and (-k,↓). In this letter, we derive a set of self-consistent equations for the correlation functions associated with the pairs that have non-opposite momentums, in which we find some new results. This means that some physics associated with superconductors should be re-explained in futures.

With all affective scatters of electrons being included in the potential $V(q)$, we take a model Hamiltonian of the form

$$H = \sum_{k,\sigma} \xi_k c^+_{k\sigma} c_{k\sigma}$$
$$+ \sum_{\substack{k,k',q \\ \sigma,\sigma'}} V(q) c^+_{k+q\sigma} c^+_{k'-q\sigma'} c_{k'\sigma'} c_{k\sigma} \quad (1)$$

The pairing is described by introducing these correlation functions

$$G(k,\sigma,\tau-\tau') = -<T_\tau c_{k\sigma}(\tau) c^+_{k\sigma}(\tau')> \quad (2)$$

$$F(k+q,\bar{k},\tau-\tau') = <T_\tau c_{k+q\sigma}(\tau) c_{\bar{k}\bar{\sigma}}(\tau')> \quad (3)$$

$$F^+(\bar{k}+q,k,\tau-\tau') = <T_\tau c^+_{\bar{k}+q\bar{\sigma}}(\tau) c^+_{k\sigma}(\tau')> \quad (4)$$

Where $\bar{k}$ represent $-k$, and $\bar{\sigma}$ represents the spin inverse to $\sigma$. When we derive a set of self-consistent equations, we must evaluate the expression with four operators. For example, in calculation we take such ways of pairing

$$\sum_{k'q\sigma'} V(q) <T_\tau c^+_{k'-q\sigma'}(\tau) c_{k'\sigma'}(\tau) c_{k-q\sigma}(\tau) c^+_{k\sigma}(\tau')>$$
$$= -\sum_{k'\sigma'} V(0) G(k',\sigma',0) G(k,\sigma,\tau-\tau')$$
$$+ \sum_q V(q) G(k-q,\sigma,0) G(k,\sigma,\tau-\tau')$$
$$- \sum_q V(q) F(k-q,\bar{k},0) F^+(\bar{k}-q,k,\tau-\tau')$$
$$- \sum_{k'} V(k) F(0,k',0) F^+(k'-k,k,\tau-\tau')$$
$$- \sum_q V(q) F(k-q,\bar{k}+q,0) F^+(\bar{k},k,\tau-\tau')$$

We will take $F(0,k',0) = 0$. After taking similar pairs above and Fourier transforming of the correlation functions, we get

$$(i\omega_n + \xi'_{\bar{k}+q}) F^+(\bar{k}+q,k,i\omega_n) =$$
$$-2\sum_{q'} V(q') F^+(\bar{k}+q-q',k+q',0) G(k,\sigma,i\omega_n)$$
$$\quad (5)$$

$$[i\omega_n - \xi'_k - \varepsilon(k,i\omega_n)] G(k,\sigma,i\omega_n) = 1 +$$
$$2\sum_q V(q) F(k-q,\bar{k}+q,0) F^+(\bar{k},k,i\omega_n) \quad (6)$$

Where $G(k,\sigma,0)$ represent $G(k,\sigma,\tau-\tau'=0)$, and so on. We also introduced

$$\xi'_k = \xi_k - 2\sum_{k'\sigma'} V(0) G(k',\sigma',0)$$
$$+ 2\sum_q V(q) G(k-q,\sigma,0)$$

$$\varepsilon(k,i\omega_n) = -4\sum_{q'\neq 0} V(q') F(k+q,\bar{k},0)$$
$$\times \frac{1}{i\omega_n + \xi'_{\bar{k}+q'}} \sum_q V(q) F^+(\bar{k}+\bar{q}-q',k+q,0)$$
$$\quad (7)$$

We can easily obtain the form of $F^+(\bar{k},k,i\omega_n)$ by (5) and (6), then obtain $G(k,\sigma,i\omega_n)$,

$F^+(\bar{k}+q,k,i\omega_n)$, $\varepsilon(k,i\omega_n)$, and so on. Let $i\omega_n \to \omega - i\delta$, we can get the self-energy of the excitations (electrons). However, we are not interested in the concrete forms of these functions in this letter, but in the related problems to superconductivity.

Firstly, compared with BCS theory, the excitation energies from the pairs with non-opposite momentum include the contribution of $\varepsilon(k,\omega)$, this is from the moves of the center of pairs. Obviously, this is more reasonable to describe the excitations in superconducting states.

Secondly, because there are pairs with non-opposite momentum, so there may be the d-p pairs in cuprate superconductors, the pairs from nearly localized holes combining with nearly free holes. These pairs may have short length pairing orders, but they should affect the properties of superconductors, such as the pseudogap behavior.

Thirdly, it is necessary to note that the excitation energies determined by (5) and (6) have the virtual parts, so the "electrons" and pairs have not the so-called long-life. It also means that the super-currents are always exchanging energies with their environments. The excitation energies in BCS theory are real, and it seems that the "pairs" are never separated off and the supper-currents are "cold and dead". In fact, the BCS theory could not explain why pairs would be separated.

Fourthly, we can find that if $F^+(\bar{k},k,i\omega_n)$ are zero, $F^+(\bar{k}+q,k,i\omega_n)$ must be zero on the basis of (5) and (6), hence the pairing temperature is the same for the pairs with opposite momentum and with non-opposite momentum. That is to say, the prediction of BCS for the superconducting temperatures of normal metals may be correct. Moreover, our results may give evidences for possible localized pairs, an electron moves around a nearly localized electron, which do not contribute to superconductivities. We must point that the energies from "separated pairs" should be decomposed into two parts, one is from the pairs with non-opposite momentum, and I think that they may be usually considered as a "background" in experiments.

In summary, consider the pairs with non-opposite momentum, we find that the excitation energies are different from the one in literatures, this must affect our knowledge on the physics of superconductors.